\documentclass[12pt]{article}
\usepackage{amsmath,amssymb,color}

\newcommand{\assign}{:=}
\newcommand{\nocomma}{}
\newcommand{\nosymbol}{}
\newcommand{\tmem}[1]{{\em #1\/}}
\newcommand{\tmop}[1]{\ensuremath{\operatorname{#1}}}
\newcommand{\tmsamp}[1]{\textsf{#1}}
\newcommand{\tmstrong}[1]{\textbf{#1}}
\newcommand{\tmtextbf}[1]{{\bfseries{#1}}}
\newcommand{\tmtexttt}[1]{{\ttfamily{#1}}}
\definecolor{grey}{rgb}{0.75,0.75,0.75}
\definecolor{orange}{rgb}{1.0,0.5,0.5}
\definecolor{brown}{rgb}{0.5,0.25,0.0}
\definecolor{pink}{rgb}{1.0,0.5,0.5}

\begin{document}

\title{Delta-gravity and Dark Energy}\author{J. Alfaro,\\
Pontificia Universidad Cat\'olica de Chile,\\
Av. Vicu\~na Mackenna 4860, Santiago, Chile,jalfaro@puc.cl\\
}\maketitle

\begin{abstract}
  We present a model of the gravitational field based on two symmetric
  tensors. The equations of motion of test particles are derived: Massive
  particles do not follow a geodesic but massless particles trajectories are
  null geodesics of an effective metric. \ Outside matter, the predictions of
  the model coincide exactly with General Relativity, so all classical tests
  are satisfied. In Cosmology, we get accelerated expansion without a
  cosmological constant.
\end{abstract}

General Relativity(GR) works very well at the macroscopic
scales{\cite{will}}. Its quantization has proved to be difficult, though. It
is non renormalizable, which prevents its unification with the other forces of
Nature. Trying to make sense of Quantum GR is the main physical motivation of
String Theories {\cite{witten}}. Moreover, recent discoveries in Cosmology
{\cite{dm,de1}} has revealed that most part of matter is in the form of
unknown matter(dark matter,DM) and that the dynamics of the expansion of the
Universe is governed by a mysterious component that accelerates the
expansion(dark energy,DE). Although GR is able to accommodate both DM and DE,
the interpretation of the dark sector in terms of fundamental theories of
elementary particles is problematic{\cite{turner}}. Although some candidates
exists that could play the role of DM, none have been detected yet. Also, an
alternative explanation based on the modification of the dynamics for small
accelerations cannot be ruled out{\cite{mond}}.

In GR, DE can be explained if a small cosmological constant($\Lambda$) is
present. At the later stages of the evolution of the Universe $\Lambda$ will
dominate the expansion, explaining the acceleration. Such small $\Lambda$ is
very difficult to generate in Quantum Field Theory (QFT) models, because in
this models $\Lambda$ is the vacuum energy, which is usually very large.

One of the most important questions in Cosmology and cosmic structure
formation is to understand the nature of dark energy in the context of a
fundamental physical theory{\cite{de2}}.

In recent years there has been various proposals to explain the observed
acceleration of the universe. They involve the inclusion of some additional
field like in quintessence, chameleon, vector dark energy or massive gravity;
Addition of higher order terms in the Einsten-Hilbert action, like f(R)
theories and Gauss-Bonnet terms; Modification of gravity on large scales by
introduction of extra dimensions. For a review, see {\cite{rde}}.

Less widely explored, but interesting possibilities, are the search for
non-trivial ultraviolet fixed points in gravity (asymptotic
safety{\cite{weias}}) and the notion of induced gravity{\cite{adler}}. The
first possibility uses exact renormalization-group techniques{\cite{litim}}
and lattice and numerical techniques such as Lorentzian triangulation
analysis{\cite{loll}}. Induced gravity proposed that gravitation is a residual
force produced by other interactions.

In a recent paper, {\cite{aep}} a two-dimensional field theory model explore
the emergence of geometry by the spontaneous symmetry breaking of a larger
symmetry where the metric is absent. Previous work in this direction can be
found in {\cite{salam}}, {\cite{ogi}} and {\cite{ru}}.\\
\ \ In this paper, we wish to present a model of gravitation that is as close
as possible to classical GR, but could make sense at the quantum level. The
main observation is that GR is finite on shell at one loop{\cite{thooft}}. In
{\cite{pedro,alfabv}} we presented a type of gauge theories, $\delta$ gauge
theories(DGT): The main properties of DGT are: 1) The classical equations of
motion are satisfied in the full Quantum theory 2) They live at one loop. 3)
They are obtained through the extension of the former symmetry of the model
introducing an extra symmetry that we call $\delta$ symmetry, since it is
formally obtained as the variation of the original symmetry. When we apply
this prescription to GR we obtain $\delta$ gravity. Quantization of $\delta$
gravity is discussed in {\cite{aag}}.

The impact of dark energy on cosmological observations can be expressed in
terms of a fluid equation of state $p = w (R) \rho$, which is to be determined
studying its influence on the large-scale structure and dynamics of the
Universe.

In this paper we follow the same approach. So we will not include the matter
dynamics, except by demanding that the energy-momentum tensor of the matter
fluid is covariantly conserved. This is required in order to respect the
symmetries of the model.

The main properties of this model at the classical level are: a)It agrees
with GR, outside the sources and with adequate boundary conditions. In
particular, the causal structure of delta gravity in vacuum is the same as in
General Relativity. So all standard test are satisfied automatically. b) When
we study the evolution of the Universe, it predicts acceleration without a
cosmological constant or additional scalar fields. The Universe ends in a Big
Rip, similar to the scenario considered in {\cite{br}}. c) The scale factor
agrees with the standard cosmology at early times and show acceleration only
at late times. Therefore we expect that density perturbations should not have
large corrections.

It should be remarked that $\delta$- gravity is not a metric model of gravity
because massive particles do not move on geodesics. Only massless particles
move on null geodesics of a linear combination of both tensor fields.

It was noticed in {\cite{pedro}} that the Hamiltonian of delta models is not
bounded from below. Phantoms cosmological models {\cite{phantom 1}},
{\cite{br}} also have this property. Although it is not clear whether this
problem will subsist in a diffeomorphism invariant model as delta gravity or
not, we mention some ways out of the difficulty at the end.

{\tmstrong{Definition of Delta gravity}} In this section we define the action
as well as the symmetries of the model and derive the equations of motion.

We use the metric convention of {\cite{weinberg}}. The action of $\delta$
gravity is:
\begin{eqnarray}
  S (g, \tilde{g}, \lambda) = \int d^d x \sqrt{- g} (- \frac{1}{2 \kappa} R +
  \mathcal{L}_M) + &  &  \nonumber\\
  \kappa_2  \int \left[ \left( R_{\mu \nu} - \frac{1}{2} g_{\mu \nu} R \right)
  + \kappa T_{\mu \nu} \right]  \sqrt{- g}  \tilde{g}^{\mu \nu} d^d x &  & 
  \label{action}+\\
  \kappa_2 \kappa \int  \sqrt{- g}  \left( \lambda^{\mu ; \nu} + \lambda^{\nu
  ; \mu} \right) T_{\mu \nu} d^d x &  &  \nonumber
\end{eqnarray}
Here $\kappa = \frac{8 \pi G}{c^4}$ , $\kappa_2$ is an arbitrary constant and
$T_{\mu \nu} \assign - \frac{2}{\sqrt{- g}}  \frac{\delta ( \sqrt{- g} 
\mathcal{L}_M)}{\delta g^{\mu \nu}}$ is the energy-momentum tensor of
matter.$R_{\mu \nu}$ is the Ricci's tensor and $R$ is the curvature scalar of
$g_{\mu \nu}$. $\tilde{g}^{\mu \nu}$ is a two- contravariant tensor under
general coordinate transformations.

The action (\ref{action}) is obtained by applying the prescription contained
in {\cite{pedro,alfabv}}. That is, we add to the action of general relativity,
the variation of it and consider the variation $\delta g_{\mu \nu} =
\tilde{g}_{\mu \nu}$ as a new field. Similarly, the symmetries we write below
are obtained as variation of the infinitesimal general coordinate
transformations where the variation of the infinitesimal parameter $\delta
\xi_0^{\rho} = \xi_1^{\rho}$ is the infinitesimal parameter of the new
transformation $\delta$. The last term in (\ref{action}) is needed to
implement the condition $T^{\mu \nu}_{; \nu} = 0$ as an equation of motion in
order to implement the $\delta$ symmetry (\ref{deltas}) off shell. This term
is not needed in vacuum.

Action (\ref{action}) is invariant under the following
transformations($\delta$):
\begin{eqnarray}
  \delta g_{\mu \nu} = g_{\mu \rho} \xi_{0, \nu}^{\rho} + g_{\nu \rho} \xi_{0,
  \mu}^{\rho} + g_{\mu \nu, \rho} \xi_0^{\rho} = \xi_{0 \mu ; \nu}^{\nosymbol}
  + \xi_{0 \nu ; \mu}^{\nosymbol} &  &  \nonumber\\
  \delta \tilde{g}_{\mu \nu} (x) = \xi_{1 \mu ; \nu}^{\nosymbol} + \xi_{1 \nu
  ; \mu}^{\nosymbol} + \tilde{g}_{\mu \rho} \xi_{0, \nu}^{\rho} +
  \tilde{g}_{\nu \rho} \xi_{0, \mu}^{\rho} + \tilde{g}_{\mu \nu, \rho}
  \xi_0^{\rho} &  &  \label{deltas}\\
  \delta \lambda_{\mu} = - \xi_{1 \mu} + \lambda_{\rho} \xi_{0, \mu}^{\rho}
   + \lambda_{\mu, \rho} \xi_0^{\rho} &  &  \nonumber
\end{eqnarray}
From now on we will fix the gauge $\lambda_{\mu} = 0$. This gauge preserves
general coordinate transformations but fixes completely the extra symmetry
with parameter $\xi_{1 \mu}$.

\tmtextbf{Equations of motion } Varying $g_{\mu \nu}$ \ we get:
\begin{eqnarray}
  S^{\gamma \sigma} + \frac{1}{2}  (R \tilde{g}^{\gamma \sigma} - g_{\mu \nu} 
  \tilde{g}^{\mu \nu} R^{\gamma \sigma}) - \frac{1}{2} g^{\gamma \sigma} 
  \frac{1}{\sqrt{- g}}  \left( \sqrt{- g} \nabla_{\nu}  \tilde{g}^{\mu \nu}
  \right)_{, \mu} + &  &  \nonumber\\
  \frac{1}{4} g^{\gamma \sigma}  \frac{1}{\sqrt{- g}}  \left( \sqrt{- g}
  g^{\alpha \beta} \nabla_{\beta} (g_{\mu \nu}  \tilde{g}^{\mu \nu})
  \right)_{, \alpha} = \kappa \frac{\delta T_{\mu \nu}}{\delta g_{\gamma
  \sigma}}  \tilde{g}^{\mu \nu} &  &  \label{s}
\end{eqnarray}
where $S^{\gamma \sigma} = (U^{\sigma \beta \gamma \rho} + U^{\gamma \beta
\sigma \rho} - U^{\sigma \gamma \beta \rho})_{; \rho \beta} \nocomma,$
$U^{\alpha \beta \gamma \rho} = \frac{1}{2}  \left[ g^{\gamma \rho} (
\tilde{g}^{\beta \alpha} - \frac{1}{2} g^{\alpha \beta} g_{\mu \nu} 
\tilde{g}^{\mu \nu}) \right]$

Varying $\tilde{g}^{\mu \nu}$ we get Einstein equation:
\begin{equation}
  \left( R_{\mu \nu} - \frac{1}{2} g_{\mu \nu} R \right) + \kappa T_{\mu \nu}
  = 0 \label{ee}
\end{equation}
Varying $\lambda_{\mu}$ we get:$T^{\mu \nu}_{; \nu} = 0$

Covariant derivatives as well as raising and lowering of indices are defined
using $g_{\mu \nu}$. Notice that outside the sources($T_{\mu \nu} = 0$), a
solution of (\ref{s}) is $\tilde{g}^{\mu \nu} = \lambda g^{\mu \nu}$, for a
constant $\lambda$, since $g^{\mu \nu}_{; \rho} = 0$ and $R_{\mu \nu} = 0$. We
will have $\tilde{g}^{\mu \nu} = g^{\mu \nu}$, assuming that both fields
satisfy the same boundary conditions far from the sources. But there exists other solutions in the vacuum. A simple case is presented in
equation (48) of {\cite{aag}}. This solution is interesting because any finite size body will 
look point-like if we watch it from far away. Studying the motion of massive and massless
test particles , using the equations we will derive below, we can see that the
parameter $\beta$ produces an additional gravitational  force. We need
further studies to ascertain whether this additional gravitational force can be used to
understand dark matter or not.

The equation for $\tilde{g}^{\mu \nu}$ is of second order in the derivatives.

{\tmstrong{Particle motion in the gravitational field}} We are aware of the presence of the gravitational field
through its effects on test particles. For this reason, here we discuss the
coupling of a test particle to a background gravitational field, such that the
action of the particle is invariant under (\ref{deltas}). 

In $\delta$ gravity we postulate the following action for a test particle:
\begin{eqnarray*}
  S_p = - m \int dt \sqrt{- g_{\mu \nu}  \dot{x}^{\mu}  \dot{x}^{\nu}} +
  \kappa_2 ' \int d^n y \sqrt{- g} \mathcal{T}_{\mu \nu} \left( \tilde{g}^{\mu \nu} +
  \lambda^{\mu ; \nu} + \lambda^{\nu ; \mu} \right)  &  & 
\end{eqnarray*}
where $\mathcal{T}_{\mu \nu}$ is the energy momentum tensor of the test particle:
\[ \mathcal{T}_{\mu \nu} (y) = \frac{m}{2 \sqrt{- g}}  \int dt \frac{\dot{x}_{\mu} 
   \dot{x}_{\nu}}{\sqrt{- g_{\alpha \beta}  \dot{x}^{\alpha} 
   \dot{x}^{\beta}}} \delta (y - x) \]
\ $\kappa_2' = \kappa_2 \kappa$ is a dimensionless constant.

That is:
\begin{eqnarray}
  S_p = m \int \frac{dt}{\sqrt{- g_{\alpha \beta}  \dot{x}^{\alpha} 
  \dot{x}^{\beta}}}  \left( g_{\mu \nu} + \frac{\kappa_2}{2}' \bar{g}_{\mu
  \nu} \right)  \dot{x}^{\mu}  \dot{x}^{\nu}  \label{geo}
\end{eqnarray}
were $\bar{g}_{\mu \nu} = \tilde{g}_{\mu \nu} + \lambda_{\mu ; \nu} +
\lambda_{\nu ; \mu}$ . Notice that
$S_p${\tmem{{\tmstrong{\tmtexttt{{\tmsamp{}}}}}}} is invariant under
(\ref{deltas}) and $t$-parametrizations.

From now on we work in the gauge $\lambda_{\mu} = 0$.

Since far from the sources, we must have free particles in Minkowski space,i.e
$g_{\mu \nu} \sim \eta_{\mu \nu}, \tilde{g}_{\mu \nu} \sim \eta_{\mu \nu}$, it
follows that we are describing the motion of a particle of mass $m' = m (1 +
\frac{\kappa_2}{2}')$

{\color{black} }Since in vacuum \ $\tilde{g}^{\mu \nu} = g^{\mu \nu}$, the
equation of motion for test particles is the same as Einstein's. Moreover, the
equation of motion \ is independent of the mass of the particle.

In order to include massless particles, we prefer to use the action
{\cite{Siegel}}:
\begin{eqnarray}
  L = \frac{1}{2}  \int dt \left( vm^2 - v^{- 1}  \left( g_{\mu \nu} +
  \kappa_2'  \bar{g}_{\mu \nu} \right)  \dot{x}^{\mu}  \dot{x}^{\nu} +
  \frac{m^2 + v^{- 2}  \left( g_{\mu \nu} + 
\kappa_2'  \bar{g}_{\mu \nu}
  \right)  \dot{x}^{\mu}  \dot{x}^{\nu}}{2 v^{- 3} g_{\alpha \beta} 
  \dot{x}^{\alpha}  \dot{x}^{\beta}} \left( m^2 + v^{- 2} g_{\lambda \rho} 
  \dot{x}^{\lambda}  \dot{x}^{\rho} \right) \right)  \label{geo3}
\end{eqnarray}
This action is invariant under reparametrizations:
\begin{equation}
  x' (t') = x (t) ; d t' v' (t') = d t v (t) ; t' = t - \varepsilon (t)
  \label{par}
\end{equation}

The equation of motion for $v$ is:
\begin{eqnarray}
  v = - \frac{\sqrt{- g_{\mu \nu}  \dot{x}^{\mu}  \dot{x}^{\nu}}}{m}  \label{v
  eq}
\end{eqnarray}
Replacing (\ref{v eq}) into (\ref{geo3}), we get back (\ref{geo}).

Let us consider first the massive case. Using (\ref{par}) we can fix the gauge
$v = 1$. Introducing $mdt = d \tau$, we get the action:
\begin{eqnarray}
  L_1 = \frac{1}{2} m \int d \tau \left( 1 - \left( g_{\mu \nu} + \kappa_2 ' 
  \bar{g}_{\mu \nu} \right)  \dot{x}^{\mu}  \dot{x}^{\nu} + \frac{1 + \left(
  g_{\mu \nu} + \kappa_2 ' \bar{g}_{\mu \nu} \right)  \dot{x}^{\mu} 
  \dot{x}^{\nu}}{2 g_{\alpha \beta}  \dot{x}^{\alpha}  \dot{x}^{\beta}} 
  \left( 1 + g_{\lambda \rho}  \dot{x}^{\lambda}  \dot{x}^{\rho} \right)
  \right)  \label{geomassive}
\end{eqnarray}
plus the constraint obtained from the equation of motion for $v$:
\begin{equation}
  g_{\mu \nu}  \dot{x}^{\mu}  \dot{x}^{\nu} = - 1 \label{shell}
\end{equation}
From $L_1$ the equation of motion for massive particles is derived. We
define:$\overline{\mathfrak{g}}_{\mu \nu} = g_{\mu \nu} + \frac{\kappa_2}{2}' 
\bar{g}_{\mu \nu}$.
\begin{eqnarray}
  \frac{d ( \dot{x}^{\mu}  \dot{x}^{\nu}  \overline{\mathfrak{g}}_{\mu \nu} 
  \dot{x}^{\beta} g_{\alpha \beta} + 2 \dot{x}^{\beta} 
  \mathfrak{\bar{g}}_{\alpha \beta})}{d \tau} - \frac{1}{2}  \dot{x}^{\mu} 
  \dot{x}^{\nu}  \mathfrak{\bar{g}}_{\mu \nu}  \dot{x}^{\beta} 
  \dot{x}^{\gamma} g_{\beta \gamma, \alpha} - \dot{x}^{\mu}  \dot{x}^{\nu} 
  \overline{\mathfrak{g}}_{\mu \nu, \alpha} = 0 &  &  \label{geo2}
\end{eqnarray}
We will discuss the motion of massive particles elsewhere.

The action for massless particles is:
\begin{eqnarray}
  L_0 = \frac{1}{4}  \int dt \left( - v^{- 1}  \left( g_{\mu \nu} + \kappa_2 '
  \bar{g}_{\mu \nu} \right)  \dot{x}^{\mu}  \dot{x}^{\nu} \right) 
  \label{massless}
\end{eqnarray}
In the gauge $v = 1$, we get:
\begin{eqnarray}
  L_0 = - \frac{1}{4}  \int dt \left( g_{\mu \nu} + \kappa_2 ' \bar{g}_{\mu
  \nu} \right)  \dot{x}^{\mu}  \dot{x}^{\nu}  \label{massless2}
\end{eqnarray}
plus the equation of motion for $v$ evaluated at $v = 1$: $\left( g_{\mu \nu}
+ \kappa_2'  \bar{g}_{\mu \nu} \right)  \dot{x}^{\mu}  \dot{x}^{\nu} = 0$

So, the massless particle moves in a null geodesic of $\mathfrak{g}_{\mu \nu}
= g_{\mu \nu} + \kappa_2'  \bar{g}_{\mu \nu}$.

{\tmstrong{Distances and time intervals}} In this section, we define the
measurement of time and distances in the model.

In GR the geodesic equation preserves the proper time of the particle along
the trajectory. \ Equation(\ref{geo2}) satisfies the same property: Along the
trajectory $\dot{x}^{\mu}  \dot{x}^{\nu} g_{\mu \nu}$ is constant.Therefore we
define proper time using the original metric $g_{\mu \nu}$,
\begin{equation}
  d \tau = \sqrt{- g_{\mu \nu} dx^{\mu} dx^{\nu}} = \sqrt{- g_{00}} dx^0 
  (dx^i = 0) \label{proper}
\end{equation}
Following {\cite{Landau}}, we consider the motion of light rays along
infinitesimally near trajectories and (\ref{proper}) to get the three
dimensional metric:
\begin{eqnarray}
  dl^2 = \gamma_{ij} dx^i dx^j, &  &  \nonumber\\
  \gamma_{ij} = \frac{g_{00}}{\mathfrak{g}_{00}}  ( \mathfrak{g}_{ij} -
  \frac{\mathfrak{g}_{0 i}  \mathfrak{g}_{0 j}}{\mathfrak{g}_{00}}) &  & 
  \label{properdistance}
\end{eqnarray}
That is, we measure proper time using the metric $g_{\mu \nu}$ but the space
geometry is determined by both metrics. In this model massive particles do not
move on geodesics of a four dimensional metric. Only massless particles move
on a null geodesic of $\mathfrak{g}_{\mu \nu}$. So, delta gravity is not a
metric theory.

{\tmstrong{The Newtonian limit}} The motion of a non relativistic particle in
a weak static gravitational field is obtained using $g_{\mu \nu} =
diag\left( - 1 - 2 \hspace{0.25em} U \epsilon, 1 - 2
\hspace{0.25em} U \epsilon, 1 - 2 \hspace{0.25em} U \epsilon, 1 - 2
\hspace{0.25em} U \epsilon \right)$, which solves Einstein equations to first
order in $\epsilon$ if $\nabla^2 U = \frac{1}{2} \kappa \rho$.

The solution for $\tilde{g}_{\mu \nu}$ \ is $\tilde{g}_{\mu \nu} =
diag\left( \epsilon \hspace{0.25em} \tilde{U}, 1 + \hspace{0.25em}
\epsilon \hspace{0.25em} \left( \tilde{U} - 2 U \right), 1 + \hspace{0.25em}
\epsilon \hspace{0.25em} \left( \tilde{U} - 2 U \right), 1 + \hspace{0.25em}
\epsilon \hspace{0.25em} \left( \tilde{U} - 2 U \right) \right)$. Solving
(\ref{s}),to first order in $\epsilon$ we get $\nabla^2  \tilde{U} =
\frac{1}{2} \kappa \rho$.

To recover the Minkowsky metric far from the sources, $\rho \rightarrow 0$, we
must require there:$U \rightarrow 0, \tilde{U} \rightarrow - \epsilon^{- 1}$.

(\ref{geo2}) implies $\frac{d^2 x^i}{dt^2} = - \phi_{, i}$ \ with $\phi = U -
\kappa_2'  (2 U + \tilde{U})$.

The Newtonian potential satisfies $\nabla^2 \phi = \frac{\kappa}{2}  (1 - 3
\kappa_2') \rho, | \kappa_2' | \ll 1$. The whole effect is a small
redefinition of Newton constant.

Gravitational red shift experiments can be used to put bounds on $\kappa_2'$.
According to (\ref{proper}), the shift in frequency of a source located at
$x_1$, compared to the same source located at $x_2$ due to the change in
gravitational potential is: $\frac{\nu_2 - \nu_1}{\nu_1} = \phi_N (x_2) -
\phi_N (x_1)$ where $\phi_N$ is the usual Newtonian potential, computed with
$\kappa$ as Newton constant. From {\cite{GPRL}} we get $\frac{\Delta \nu}{\nu}
= (1 + 2.5 \pm 70 \times 10^{- 6})  (\varphi_S - \varphi_E + \ldots .)$, where
$\varphi_S$ is the gravitational potential at the spacecraft position and
$\varphi_E$ is the gravitational potential on Earth. $\ldots$ accounts for
additional effects not related to the gravitational potential. We can ascribe
the uncertainty of the experiment to $\kappa_2'$, to get the bound:
\[ \left. \left| \kappa_2' \right| < 24 \times 10^{- 6} \right. \]
This bound is conservative because the Newton constant itself has a larger
error {\cite{CODATA}}: $G = 6.67428 \pm 0.00067 \times 10^{- 11} 
\frac{m^3}{\tmop{kgs}^2}$

In our description of the evolution of the Universe, the value of $\kappa_2'$
is not important, so we will keep it arbitrary for the time being.

{\tmstrong{Friedman-Robertson-Walker(FRW) metric}} This is the main section of
the paper. We discuss the equations of motion for the Universe described by
the FRW metric. We use  spatial curvature equal to zero to agree with cosmological observations.

In this paper we will deal only with a perfect fluid, since rotational and
translational invariance implies that the energy-momentum tensor of the
Universe has this form.The energy momentum tensor for a perfect fluid is
{\cite{weinberg}}:
\begin{equation}
  T_{\mu \nu} = pg_{\mu \nu} + (p + \rho) U_{\mu} U_{\nu}, g^{\mu \nu} U_{\mu}
  U_{\nu} = - 1
\end{equation}
Then:
\begin{equation}
  \frac{\delta T_{\mu \nu}}{\delta g_{\gamma \sigma}}  \tilde{g}^{\mu \nu} = p
  \tilde{g}^{\gamma \sigma} + \frac{1}{2}  (p + \rho)  (U^{\gamma} U_{\nu} 
  \tilde{g}^{\sigma \nu} + U^{\sigma} U_{\nu}  \tilde{g}^{\gamma \nu})
  \label{tfluid}
\end{equation}
In this case, assuming flat three dimensional metric:
\begin{eqnarray*}
  - ds^2 = dt^2 - R (t)^2  \left\{ dr^2 + r^2 d \theta^2 + r^2 \sin^2 \theta d
  \phi^2 \right\} &  & \\
  - d \tilde{s}^2 = \tilde{A} (t) dt^2 - \tilde{B} (t)  \left\{ dr^2 + r^2 d
  \theta^2 + r^2 \sin^2 \theta d \phi^2 \right\} &  & 
\end{eqnarray*}
Using (\ref{geo2}, \ref{proper}), we can check that these are co-mobile
coordinates and the proper time interval $d \tau$ for a co-moving clock is
just $dt$, so $t$ is the time measured in the rest frame of a co-moving clock.
Equations (\ref{s}, \ref{tfluid}) give:
\begin{eqnarray}
  - \dot{R}  \dot{\tilde{B}} - \frac{1}{2} pR \tilde{B} + \frac{1}{2} R^{- 1} 
  \dot{R}^2  \tilde{B} - \frac{1}{6} \rho R^3  \tilde{A} + \frac{3}{2} R
  \dot{R}^2  \tilde{A} = 0 &  &  \nonumber\\
  - p \tilde{B} - 2 \ddot{\tilde{B}} - R^{- 2}  \dot{R}^2  \tilde{B} + 2 R^{-
  1}  \ddot{R}  \tilde{B} + 2 R^{- 1}  \dot{R}  \dot{\tilde{B}} + &  & 
  \nonumber\\
  \rho \hspace{0.25em} R^2  \tilde{A} + \dot{R}^2  \tilde{A} + 2 R \dot{R} 
  \dot{\tilde{A}} + 2 R \tilde{A}  \ddot{R} = 0 &  &  \label{rw4}
\end{eqnarray}
Einstein's equations are:
\begin{eqnarray*}
  \frac{3 \hspace{0.25em} \left( \frac{d}{d \hspace{0.25em} t} 
  \hspace{0.25em} R \right)^2}{R^2} = \kappa \rho &  & , 2 \hspace{0.25em} R
  \hspace{0.25em} \left( \frac{d^2}{d \hspace{0.25em} t^2}  \hspace{0.25em} R
  \right) + \left( \frac{d}{d \hspace{0.25em} t}  \hspace{0.25em} R \right)^2
  = - \kappa R^2 p\\
  &  & 
\end{eqnarray*}
We use the equation of state $p = w \rho$, to get, for $w \neq - 1$ :
\begin{eqnarray}
  R = R_0 t^{\frac{2}{3 (1 + w)}}, \tilde{A} = 3 wl_2 t^{( \frac{w - 1}{w +
  1})}, \nonumber\\
  \tilde{B} = R_0^2 l_2 t^b, b = \frac{4}{3 w + 3} + \frac{w - 1}{w + 1} 
  \label{rw}
\end{eqnarray}
$l_2$ is a free parameter.

{\tmstrong{Red Shift}} To make the usual connection between redshift and the
scale factor, we consider light waves \ traveling to \ $r = 0$, from $r =
r_1$, along the $r$ direction with fixed $\theta, \phi$. Photons moves on a
null geodesic of $\mathfrak{g}$:
\begin{equation}
  0 = - (1 + \kappa_2'  \tilde{A}) dt^2 + (R^2 + \kappa_2'  \tilde{B})  (dr^2
  + r^2 d \theta^2 + r^2 \sin^2 \theta d \phi^2)
\end{equation}
So,
\begin{equation}
  \int_{t_1}^{t_0} dt_{\nosymbol}  \sqrt{\frac{1 + \kappa_2' tA}{R^2 +
  \kappa_2' tB}} = r_1 \label{r1}
\end{equation}
A typical galaxy will have fixed \ $r_1, \theta_1, \phi_1$. If a second wave
crest is emitted at $t = t_1 + \delta t_1$ from $r = r_1$, it will reach $r =
0$ at $t_0 + \delta t_0$, where
\[ \int_{t_1 + \delta t_1}^{t_0 + \delta t_0} dt_{\nosymbol}  \sqrt{\frac{1 +
   \kappa_2' tA}{R^2 + \kappa_2' tB}} = r_1 \]
Therefore, for $\delta t_1, \delta t_0$ small, which is appropiate for light
waves, we have:
\begin{equation}
  \delta t_0  \sqrt{\frac{1 + \kappa_2' tA}{R^2 + \kappa_2' tB}} (t_0) =
  \delta t_1  \sqrt{\frac{1 + \kappa_2' tA}{R^2 + \kappa_2' tB}} (t_1)
\end{equation}
Introduce:
\[ \tilde{R} (t) = \sqrt{\frac{R^2 + \kappa_2' tB}{1 + \kappa_2' tA}} (t) \]
We get:$\frac{\delta t_0}{\delta t_1} = \frac{\tilde{R} (t_0)}{\tilde{R}
(t_1)}$ . A crucial point is that, according to equation (\ref{proper}),
$\delta t$ measure the change in proper time. That is:$\frac{\nu_1}{\nu_0} =
\frac{\tilde{R} (t_0)}{\tilde{R} (t_1)}$, where $\nu_0$ is the light frequency
detected at $r = 0$ corresponding to a source emission at frequency $\nu_1$.
Or in terms of the redshift parameter $z$, defined as the fractional increase
of the \ wavelength $\lambda$:
\begin{equation}
  z = \frac{\tilde{R} (t_0)}{\tilde{R} (t_1)} - 1 = \frac{\lambda_0 -
  \lambda_1}{\lambda_1}
\end{equation}
We see that \ $\tilde{R} $ replaces the usual scale factor $R$ in the
computation of $z$.

\tmtextbf{Luminosity distance} Let us consider a mirror of radius $b$ that is
receiving light from a distant source. The photons that reach the mirror are
inside a cone of half-angle $\varepsilon$ with origin at the source.

Let us compute $\varepsilon$.The light path of rays coming from a far away
source at $\vec{x}_1$ is given by $\vec{x} \left( \rho \right) = \rho \hat{n}
+ \vec{x}_1$, $\rho > 0$ is a parameter and $\hat{n}$ is the direction of the
light ray.The path reaches us at $\vec{x} = 0$ for $\rho = \left| \vec{x}_1
\right| = r_1$. So $\hat{n} = - \hat{x}_1 + \vec{\varepsilon}$. Since
$\hat{n}, \hat{x}_1$ have modulus 1,  $\varepsilon = \left| \vec{\varepsilon}
\right|<<1$ is precisely the angle between $- \vec{x}_1$ and $\hat{n}$ at the
source.The impact parameter is the proper distance of the path from the
origin, when  $\rho = \left| \vec{x}_1 \right|$. The proper distance is
determined by the 3-dimensional metric (\ref{properdistance}). That is $b =
\tilde{R} \left( t_0 \right) r_1 \theta = \tilde{R} \left( t_0 \right) r_1
\varepsilon$, i.e. $\varepsilon = \frac{b}{\tilde{R} \left( t_0 \right) r_1}$.

Then the  solid angle of the cone is $\pi \varepsilon^2 = \frac{A}{r_1^2
\tilde{R} \left( t_0 \right)^2}$, where $A = \pi b^2$ is the proper area of
the mirror.The fraction of all isotropically emitted photons that reach the
mirror is $f = \frac{A}{4 \pi r_1^2 \tilde{R} \left( t_0 \right)^2}$. Each
photon carries an energy $h \nu_1$ at the source and \ $h \nu_0$ at the
mirror. Photons emitted at intervals $\delta t_1$ will arrive at intervals
$\delta t_0$. We have $\frac{\nu_1}{\nu_0} = \frac{\tilde{R} (t_0)}{\tilde{R}
(t_1)}, \frac{\delta t_0}{\delta t_1} = \frac{\tilde{R} (t_0)}{\tilde{R}
(t_1)}$. Therefore the power at the mirror is $P_0 = L \frac{\tilde{R}
(t_1)^2}{\tilde{R} (t_0)^2} f$, where $L$ is the luminosity of the
source. The apparent luminosity is $l = \frac{P_0}{A} = L \frac{\tilde{R}
(t_1)^2}{\tilde{R} (t_0)^2} \frac{1}{4 \pi r_1^2 \tilde{R} \left( t_0
\right)^2}$. In Euclidean space, the luminosity decreases with distance $d$
according to $l = \frac{L}{4 \pi d^2}$.This permits to define the luminosity
distance:$d_L = \sqrt{\frac{L}{4 \pi l}} = \tilde{R} (t_0)^2
\frac{r_1}{\tilde{R} (t_1)}$. Using (\ref{r1}) we can write this in terms of
the red shift:
\begin{equation}
  d_L = (1 + z)  \int_0^z \frac{dz'}{\tilde{H} (z')}, \tilde{H} =
  \frac{\dot{\tilde{R}}}{\tilde{R}}
\end{equation}
\tmtextbf{Supernova Ia data}

 The supernova Ia data gives, $m$ (apparent or effective magnitude) as a
function of $z$. This is related to distance $d_L$ by $m = M + 5 log (
\frac{d_L}{10 pc})$. Here $M$ is common to all supernova and $m$ changes with
$d_L$ alone.

We compare $\delta$ gravity to General Relativity(GR) \ with a cosmological
constant:
\[ H^2 = H_0^2  (\Omega_m (1 + z)^3 + (1 - \Omega_m)), \Omega_{\Lambda} = 1 -
   \Omega_m \]
Notice that $\tilde{A} = 0$ for $w = 0$ in (\ref{rw}). So, it seems that we
cannot fit the supernova data. However $w = 0$ is not the only component of
the Universe. The massless particles that decoupled earlier still remain. It
means that the true $w$ is between \ $0 \leqslant w < \frac{1}{3}$ , but very
close to $w = 0$. So, we will fit the data with $w = 0.1, 0.01, 0.001$ and see
how sensitive the predictions are to the value of $w$.

Using data from Essence{\cite{essence}}, we notice that $R^2$ test changes
very little for the chosen sequence of $w$'s. Each fit determines the best
$l_2$ for a given $w$. In this way we see that $l_2$ scales like $l_2 \sim
\frac{a}{3 w}$, $a$ being independent of $w$. As an approximation to the limit
$w = 0$, we get:
\begin{equation}
  \tilde{R} (t) = R (t) \frac{\sqrt{a}}{\sqrt{a - t}} \label{rip}
\end{equation}
$\sqrt{\frac{1}{3 w}}$renormalizes the derivative of $\tilde{R}$ at $t = 0$.
It is not divergent, because for $t \rightarrow 0$, $w \rightarrow
\frac{1}{3}$. $a$ is a free parameter determined by the best fit to the data.

Of course, the complete model must include the contribution of normal
matter($w = 0$) plus relativistic matter $(w = \frac{1}{3})$. But, at later
times, the data should tend to (\ref{rip}).

Let us fit the data to the simple scaling model (\ref{rip}).

We get:

$\Omega_m = 0.22 \pm 0.03, M = 43.29 \pm 0.03$ , $\chi^2  (perpoint) =
1.0328$, General Relativity

$a = 2.21 \pm 0.12, M = 43.45 \pm 0.06$, $\chi^2  (perpoint) = 1.0327$, Delta
Gravity

$\delta$-gravity with non-relativistic(NR) matter alone give a fit to the data
as good as GR with NR matter plus a cosmological constant.

According to the fit to data, a Big Rip will happen at $t = 2.21049$ in
unities of $t_0$(today). It is a similar scenario as in {\cite{br}}.

Finally, we want to point out that since for $t \rightarrow 0$, we have $w
\rightarrow \frac{1}{3}$, then $\tilde{R} (t) = R (t)$. Therefore the
accelerated expansion is slower than (\ref{rip}) when we include both matter
and radiation in the model.

{\tmstrong{Conclusions and Open Problems}} \ Delta Gravity agrees with General
Relativity when $T_{\mu \nu} = 0$, imposing same boundary conditions for both
tensor fields. In particular, the causal structure of delta gravity in vacuum
is the same as in General Relativity, since in this case the action \
(\ref{geo}) \ is proportional to the geodesic action in GR.

We recover the Newtonian approximation.

In a homogeneous and isotropic universe, we get accelerated expansion without
a cosmological constant or additional scalar fields.

The computation of PPN(Postnewtonian) parameters is in progress, but we do not
expect large departures from general relativity, because the newtonian limit
is the right one, as explained in section 6. Moreover the interestellar space
has very small matter densities, so $\delta$- gravity must give general
relativity values for the PPN parameters(See comments after
equation(\ref{ee})). Additionally, please notice that all $\tilde{g}$
contributions are multiplied by the small parameter $\kappa_2'$ of the order
of $10^{- 5}$ or less, so they are much suppressed in the solar system.

Stellar evolution will not be changed from its newtonian description, unless
density of matter becomes very large. Even at the densities of white dwarfs
the Poisson equation for the gravitational potential suffices.(See, for
instance {\cite{weinberg}}, chapter 11.3). \ $\delta$- gravity implies it, as
it is shown in section 6. Higher densities which are present in neutron stars
may provide new tests of \ \ $\delta$-gravity, since there we have to use the
whole non-linear Einstein equations and the corresponding \ $\delta$- gravity
equations. But for the inner regions of massive stars, data is very scarce.

Notice that equation (\ref{rw}) implies that \ $\tilde{R} = R$ at the
beginning of the Universe, when $w = \frac{1}{3}$, corresponding to
ultrarelativistic matter. That is, the accelerated expansion started at a
later time, which is needed if we want to recover the observational data of
density perturbations and growth of structures in the Universe. An earlier
acceleration of the expansion would prevent the growth of density
perturbations.

Work is in progress to compute the growth of density perturbations, \ the
anisotropies in the CMBR, BAO, WL and the evolution of massive stars. The
comparison of these calculations with the considerable amount of astronomical
data that will be available in the near future will be a very stringent test
of the present gravitational model.

It was noticed in {\cite{pedro}} that the Hamiltonian of delta models is not
bounded from below. Phantoms cosmological models {\cite{phantom 1}},
{\cite{br}} also have this property. Although it is not clear whether this
problem will subsist in a diffeomorphism invariant model as delta gravity or
not, we want to mention some ways out of the difficulty.

a) Delta gravity is a gauge theory. Moreover it is diffeomorphism invariant.
Thus the canonical Hamiltonian vanishes identically. It may be possible to
truncate the Hilbert space, using the BRST formalism, to define a model with a
Hamiltonian bounded from below. This is a difficult task that goes far beyond
the present paper, but should be pursued in a future work.

b) In a supersymmetric model we have $H = Q^2$, where $H$ is the Hamiltonian
and $Q$ is the hermitian supersymmetry charge. Thus the Hamiltonian is bounded
from below. So, we expect that a delta supergravity model has a Hamiltonian
bounded from below.

{\tmstrong{Acknowledgements}} The work of JA is partially supported by
VRAID/DID/46/2010 and Fondecyt 1110378. He wants to thank R. Avila and P.
Gonz\'alez for several useful remarks; The author acknowledges interesting
conversations with L. Infante, G. Palma, M. Ba\~nados and A. Clocchiatti. In
particular, JA wants to thank A. Clocchiatti for pointing out the data in
{\cite{essence}}. Finally, JA wants to thank J. Gamboa for a careful reading
of the manuscript.


\begin{thebibliography}{99}
  \bibitem{will}Clifford M. Will, "The Confrontation between General
  Relativity and Experiment", Living Rev. Relativity 9, (2006),
  http://www.livingreviews.org/lrr-2006-3;Slava G. Turyshev, Annual Review of
  Nuclear and Particle Science, Vol. 58: 207-248 (Volume publication date
  November 2008)
  
  \bibitem{witten}For a modern review of string model see: M.B. Green, J.H.
  Schwarz and E. Witten, "Superstring Theory ", vols. 1, 2, Cambridge
  University Press 1987. J. Polchinski, "String Theory", vols. 1,2, Cambridge
  University Press 1998.
  
  \bibitem{dm}for a review of Dark Matter and its detection , see S.
  Weinberg, Cosmology, Oxford University Press 2008; Hooper, D. and Baltz,
  Annu. Rev. Nucl. Part. Sci. 58, 293314(2008).
  
  \bibitem{de1}A. G. Riess et al. (Supernova Search Team), Astron. J. 116,
  1009 (1998),S. Perlmutter et al. (Supernova Cosmology Project), Astrophys.
  J. 517, 565 (1999). For a recent review, see R. R. Caldwell and M.
  Kamionkowski, The Physics of Cosmic Acceleration,astro-ph 0903.0866.
  
  \bibitem{turner}Frieman, J. A., Turner, M. S. \& Huterer, D. Dark energy
  and the accelerating Universe. Annu. Rev. Astron. Astrophys. 46, 385432
  (2008).
  
  \bibitem{mond}Milgrom, M.,A modification of the Newtonian dynamics as a
  possible alternative to the hidden mass hypothesis, ApJ, 270, 365(1983);
  Bekenstein, J. Relativistic gravitation theory for the MOND paradigm, Phys.
  Rev. D 70, 083509 (2004)
  
  \bibitem{rde}See, for instance, S. Tsujikawa, Lect.Notes
  Phys.800:99-145,2010
  
  \bibitem{weinberg}S. Weinberg, Gravitation and Cosmology(Wiley, New York,
  1972)
  
  \bibitem{weias}S. Weinberg, in General Relativity: An Einstein centenary
  survey, edited by S. W.Hawking and W.Israel (Cambridge University Press,
  1979), chapter 16, p. 790.
  
  \bibitem{adler}Ya.B. Zeldovich, JETP Lett., 6, 316 (1967); A. Sakharov,
  SOv. Phys. Dokl., 12, 1040 (1968); O. Klein, Phys. Scr. 9, 69 (1974); S.
  Adler, Rev. Mod. Phys., 54, 729 (1982).
  
  \bibitem{litim}D.F. Litim, Phys.Rev.Lett.92:201301,2004; AIP Conf. Proc.
  841, 322 (2006); e-Print: arXiv:0810.367; A. Codello, R. Percacci and C.
  Rahmede, Annals Phys.324:414-469,2009; M. Reuter and F. Saueressig, Lectures
  given at First Quantum Geometry and Quantum Gravity School, Zakopane, Poland
  (2007),arXiv:0708.1317
  
  \bibitem{loll}J. Ambjorn, J. Jurkiewicz and R. Loll,
  Phys.Rev.Lett.85:924-927,2000.
  
  \bibitem{aep}J. Alfaro, D. Espriu and D. Puigdomenech, Phys. Rev.
  D82:045018,2010.
  
  \bibitem{salam}C.J. Isham, A. Salam and J.A. Strathdee, Annals
  Phys.62:98-119,1971.
  
  \bibitem{ogi}A.B. Borisov and V.I. Ogievetsky,
  Theor.Math.Phys.21:1179,1975; E.A. Ivanov and V.I. Ogievetsky,
  Lett.Math.Phys.1:309-313,1976.
  
  \bibitem{ru}D. Amati and J. Russo, Phys.Lett. B 248, 44 (1990); J. Russo,
  Phys.Lett. B 254, 61 (1991); A. Hebecker, C. Wetterich,
  Phys.Lett.B574:269-275,2003; C. Wetterich, Phys. Rev. D 70: 105004, 2004.
  
  \bibitem{de2}See, for instance:Albrecht, A., et al., 2006,
  astro-ph/0609591 and Peacock. J.A., et al., 2006, astro-ph/0610906
  
  \bibitem{thooft}G. 't Hooft and M. Veltman, Ann. Inst. Henri Poincar, 20
  (1974) 69
  
  \bibitem{alfabv}J. Alfaro, bv gauge theories, hep-th 9702060.
  
  \bibitem{pedro}J. Alfaro and P. Labra\~na,Phys. Rev. D 65, 045002 (2002).
  
  \bibitem{aag} J. Alfaro, P. Gonzalez, R. Avila, Class.Quant.Grav. 28
  (2011) 215020.
  
  \bibitem{phantom 1}R.R. Caldwell, Physics Letters B 545 (2002) 2329.
  
  \bibitem{br}R. R. Caldwell, M. Kamionkowski, and N. N.Weinberg, Phys. Rev.
  Lett.91(2003)071301.
  
  \bibitem{Siegel}W. Siegel, "Fields",hep-th 9912205v3, page 193.
  
  \bibitem{Landau}L. Landau and L.M. Lifshitz, The Classical Theory of
  Fields, Pergamon Press 1980.
    
  \bibitem{GPRL}R. F. C. Vessot, et al, Phys. Rev. Lett. 45, 2081 (1980).
  
  \bibitem{CODATA}Mohr, Peter J.; Taylor, Barry N.; Newell, David B. , Rev.
  Mod. Phys. 80: 633730.
  
  \bibitem{essence}W. M. Wood-Vasey et al.,Astrophys.J.666:694-715,2007.
\end{thebibliography}
\end{document}